\documentclass[10pt]{iopart1}
\usepackage{iopams,bbm}  
\begin{document}

\letter{Ballistic transport in random magnetic fields 
  with anisotropic long-ranged correlations}

\author{Hajo Leschke\dag, 
        Simone Warzel\ddag\ \footnote[3]{To
whom correspondence should be addressed (swarzel@princeton.edu). On leave from: Institut f\"ur Theoretische Physik, Universit\"at Erlangen-N\"urnberg, 
Staudtstrasse 7, 91058 Erlangen, Germany}
        and \\ Alexandra Weichlein\dag\ }

\address{\dag\ Institut f\"ur Theoretische Physik, Universit\"at Erlangen-N\"urnberg, 
Staudtstrasse 7, 91058 Erlangen, Germany}

\address{\ddag\ Jadwin Hall, Princeton University, Princeton, NJ 08544, USA}

\begin{abstract}  We present exact theoretical results about energetic and dynamic properties of 
  a spinless charged quantum particle on the Euclidean plane 
  subjected to a perpendicular random magnetic field of Gaussian type with non-zero mean.  
  Our results refer to the simplifying but remarkably illuminating limiting case
  of an infinite correlation length along one direction and a finite but strictly positive correlation length along the 
  perpendicular direction in the plane. They are therefore ``random analogs'' of results first obtained by 
  A.\ Iwatsuka in 1985 and by J.\ E.\ M\"uller in 1992, which are greatly esteemed, in particular for providing a basic understanding 
  of transport properties in certain quasi-two-dimensional semiconductor heterostructures subjected to non-random 
        inhomogeneous magnetic fields.

\end{abstract}
\pacs{72.15.Gd, 72.20.My, 73.23.Ad, 75.47.Jn}


\maketitle

Quantum-mechanical models for a single spinless electrically charged particle
on the (infinitely extended) Euclidean plane $ \mathbb{R}^2 $ subjected to a perpendicular
spatially random magnetic field (RMF) have become 
a topic of growing interest over the last decade. 
Such models are currently discussed in relation with magneto-transport properties 
of quasi-two-dimensional semiconductor heterostructures 
with certain randomly built-in magnets \cite{GBGB94,Smi94,Man95,AAKI00,Rus00,Byk01,Rus03}. 
Moreover, they are part of effective theories 
for the fractional quantum Hall effect \cite{Hei98,Wol00,MuSh03}. 
Just like in Anderson's model \cite{And58} of a quantum
particle in a random scalar potential (and no or a constant magnetic field)\footnote{For a recent survey of rigorous results in case of continuum models see \cite{LMW03}.}, 
the fundamental question is to understand the
spectral and transport properties of the underlying Hilbert-space operator 
representing the (kinetic) energy and generating the dynamics of the particle in a RMF.
Until recently, different studies by perturbative,
quasi-classical, field-theoretical, and numerical methods have given partially
conflicting answers \cite{ArMi94,LeCh94,KaOh95,ShWe95,YaBa96,BaSc98,Eve99,Fur99,PoSc99,KaKr99,ShWe00,Yak00,TaEf00,TaEf01,Ngu02,KaOh03,EfKo03}.

Since ``the power and utility of simple models can hardly be overestimated'' \cite{Kit66}, 
the purpose of this Letter is to present first exact 
(de)localization results in case of simple, 
but remarkably illuminating RMF's\footnote{Ref.~\cite{Klopp}
outlines a rigorous proof of the existence of localized states at low energies for certain RMF's on the (infinite) square lattice $ \mathbb{Z}^2 $
instead of the two-dimensional continuum $ \mathbb{R}^2 $.}. The simplification arises from the assumption that the
fluctuations of the RMF on $ \mathbb{R}^2 = \mathbb{R} \times \mathbb{R} $ are anisotropically
long-ranged correlated in the sense that we consider the limiting case
of an infinite correlation length along one direction and take the
correlation length to be finite but strictly positive along the
perpendicular direction in the plane.  In other words, we assume the RMF to be independent of one of the two Cartesian co-ordinates, which
we choose to be the second one, $ x_2 $.  The remaining dependence of
the RMF-values on the first
co-ordinate $ x_1 $ we suppose to be governed by the realizations
$ b:=\{ b(x_1) \}_{x_1 \in \mathbb{R}} $ 
of a homogeneous and \emph{ergodic} real-valued 
random (or: stochastic) process with the real line $ \mathbb{R} = ]-\infty ,\infty[$ as its parameter set \cite{CrLe67}. 
We will assume throughout that its mean $ \overline{b(0)} $ is non-zero and
finite, 
\begin{equation}\label{eq:mean} 0 < \big| \, \overline{b(0)} \, \big| < \infty.
\end{equation} 
Here the overbar denotes the probabilistic (or: ensemble) average. 
Taking the (Lebesgue-)integral $ a_2(x_1) := \int_0^{x_1} dx_1' b(x_1') $, which exists almost surely for all $ x_1 \in \mathbb{R} $, 
as the second component of 
the vector potential $ ( 0, a_2(x_1) ) $ 
in the asymmetric gauge, the Hamiltonian (or: kinetic-energy operator) is then given as
\begin{equation} \label{def:H}
H := \frac{1}{2} \left[ P_1^2 + ( P_2 - a_2(Q_1))^2 \right]
\end{equation}
in terms of the two
components of the usual canonical momentum and position operators, $ P_1 $, $ P_2 $, respectively $ Q_1 $, ($ Q_2 $,)
corresponding to the $ x_1 $- and the $ x_2 $-direction.  
All operators act self-adjointly on 
the Hilbert space $ {\rm L}^2(\mathbb{R}^2) = {\rm L}^2(\mathbb{R}) \otimes {\rm L}^2(\mathbb{R}) $ of 
square-integrable, complex-valued  functions on the plane $\mathbb{R}^2$.
For notational transparency we use physical units such that Planck's constant (divided by $ 2 \pi $) and the particle's mass and charge 
are all equal to $ 1 $.\\

\noindent {{\it Energetic properties.}~The nice feature of $ H $ is its translational invariance along the 
$ x_2 $-direction so that it commutes with $ P_2 $, an operator which can be partially Fourier
decomposed on  $ {\rm L}^2(\mathbb{R}^2)$ 
according to $ P_2 = \int_{-\infty}^\infty dk \, k \, \mathbbm{1} \otimes | k \rangle \langle k | $ (using an informal notation). 
Therefore $ H $ can be decomposed according to 
\begin{equation}\label{eq:decomp}  H = \int_{-\infty}^\infty dk\,  H(k) \otimes | k \rangle \langle k | 
\end{equation}
into the one-parameter family 
 \begin{equation}\label{eq:redH}
  H(k) := \frac{1}{2} \big[ P_1^2 + ( k \, \mathbbm{1}  - a_2(Q_1) )^2 \big], \quad k \in \mathbb{R},
\end{equation}
of effective (or: fibre) Hamiltonians on the Hilbert space $ {\rm L}^2(\mathbb{R}) $ for the one-dimensional motion along the $ x_1 $-direction.
Here each wave number $ k \in \mathbb{R} $ is a possible (spectral) value of the particle's canonical momentum along the $ x_2  $-direction. 
For any typical realization $ b $ the Birkhoff-Khinchin ergodic theorem \cite{CrLe67,CoFoSi82}, 
\begin{equation}\label{eq:erg}
  \lim_{| x_1 | \to \infty} \frac{a_2(x_1)}{ x_1} = \overline{b(0)} \neq 0, 
\end{equation}
ensures that the effective scalar potential entering $ H(k) $ 
confines the particle along the $ x_1 $-direction for large $ | x_1 | $ quadratically. As a consequence, 
for each fixed $ k \in \mathbb{R} $ the operator $ H(k) $ has purely discrete spectrum 
with strictly positive and non-degenerate eigenvalues 
$ 0 < \varepsilon_0(k) <  \varepsilon_1(k) < \ldots  $ so that its spectral resolution reads 
\begin{equation}\label{eq:redHs}
  H(k) = \sum_{n =0}^\infty \varepsilon_n(k) \, | \varphi_n(k) \rangle \langle \varphi_n(k) | 
\end{equation}
with 
normalized and pairwise orthogonal eigenfunctions $ | \varphi_0(k) \rangle,  | \varphi_1(k) \rangle , \ldots  $ 
spanning $ {\rm L}^2(\mathbb{R}) $. 
By (\ref{eq:decomp}) and (\ref{eq:redHs}) the spectrum of $ H $ is given by a set-theoretic union,
\begin{equation}
  \mathrm{spec} \, H = \bigcup_{n = 0 }^\infty \beta_n,\quad 
  \beta_n := \Big[ \inf_{k \in \mathbb{R}} \varepsilon_n(k) , \sup_{k \in \mathbb{R}} \varepsilon_n(k) \Big].
\end{equation}
Here the closed interval
$  \beta_n $ 
is the $ n $th energy band. It is a subset of the positive half-line $ [0,\infty[ $ and extends from the
lower to the upper edge of the $ n $th energy-band function $ \varepsilon_n $.
A further important consequence of the assumed ergodicity is that, although the spectrum of $ H(k) $ for fixed $ k \in \mathbb{R} $ 
in general depends on $ b $, 
each resulting energy band $ \beta_n $ of $ H $ is non-random almost surely, that is, the same for all typical $ b $.  

The random Hamiltonian $ H $ with the non-random energy-band structure of its spectrum 
is a random variant of models first investigated in \cite{Iwa85} 
and (non-rigorously) in the often quoted paper \cite{Mul92}. 
By studying special non-random $ b $'s 
these and other works \cite{Kra95,MaPu97,RePe00,NoBe92,Sim01,Law02,LeRu02} have illustrated
that a non-constant $ b $  has a tendency to delocalize the particle 
along the $ x_2 $-direction. In fact, according to classical mechanics 
a particle with non-zero kinetic energy wanders off to infinity along snake or cycloid-like orbits 
winding around (straight) contours of constant magnetic field \cite{CyFr87,Mul92}. 
The quantum analog of this unbounded motion should manifest itself in the exclusive appearance of 
(absolutely) continuous spectrum of $ H $, or equivalently, of only strictly positive bandwidths,   
$ | \beta_n | := \sup_{k \in \mathbb{R}} \varepsilon_n(k) - \inf_{k \in \mathbb{R}} \varepsilon_n(k) > 0 $ 
for all $ n $. 
While plausible from the (quasi{-)}classical picture, 
the rigorous exclusion of flat energy bands is not trivial and has been accomplished 
so far only for certain classes of non-constant\footnote{In the constant case, that is, $ b(x_1) = b_0 $ for all $ x_1 \in \mathbb{R} $ 
with some constant $ b_0 \neq 0 $, each eigenvalue of $ H(k) $ is independent of $ k \in \mathbb{R} $ and given by  
a Landau level \cite{Foc28,Lan30}, $ \varepsilon_n(k) = (n + 1/2) | b_0 | $, so that $ | \beta_n | = 0 $ for all $ n $.} 
but non-random $ b $'s \cite{Iwa85,MaPu97}.
Our main theorem establishes for the first time such a result in the random case.

\vspace*{0.1cm}
\noindent
{{THEOREM.}~~{\it
If the RMF is given by a homogeneous Gaussian random process with its 
mean $ \overline{b(0)} $ obeying (\ref{eq:mean}) and its covariance function 
\begin{equation}
  c(x_1) := \overline{b(x_1)\, b(0) } - \big(\overline{b(0)}\big)^2 
\end{equation}
fulfilling the following two requirements:\\[1ex]
  \indent (i)~~ $ c $ is continuous at the origin (and hence everywhere) with $ 0 < c(0) < \infty $, 
  
  \indent (ii)~~ $ \lim_{\ell \to \infty} \ell^{-1} \int_0^\ell d x_1 \, \big(c(x_1)\big)^2 = 0 $, \\[1ex]
\noindent then $ | \beta_n | > 0 $ for all energy-band indices $ n $ and $ \mathrm{spec} \, H = [0, \infty[ $, almost surely.}

\vspace*{0.1cm}
Given our simplifying a-priori assumption, the two requirements are both mathematically mild and physically relevant. 
By the first one the RMF is neither non-random nor delta-correlated and has realizations which are continuous in the mean-square sense. 
Because of the Bochner-Khinchin \cite{CrLe67,CoFoSi82}, the Fomin-Grenander-Maruyama \cite{CrLe67,CoFoSi82}, and the Wiener theorem \cite{CoFoSi82,CyFr87}, the 
second requirement is then equivalent to the ergodicity of the underlying Gaussian random process. 
In particular, (ii) requires that the correlation of the RMF's fluctuations at two different points in the plane exhibits some decay 
with increasing absolute difference of their first co-ordinates. 
The simple condition $ \lim_{|x_1|\to \infty} c(x_1) = 0 $ is sufficient but not necessary.

The basic observation for the proof of the theorem is that (i) and (ii) ensure a non-zero probability for the 
occurrence of realizations $ b $ with arbitrarily small absolute values on spatial average over arbitrarily long line segments, 
that is
\begin{equation}\label{eq:topol}
  \mathrm{Prob}\left\{ \int_{-\ell}^{\ell} dx_1 \, | b(x_1) | < \delta \right\} > 0  
\end{equation}
for all $ \ell > 0 $ and all $ \delta > 0 $.  
Such realizations, although rare because of $ \overline{b(0)} \neq 0 $, come with nearly free motion. 
More precisely, for any given (arbitrarily small) energy $ \varepsilon > 0 $ and any (arbitrarily large) integer $ n_0 \geq 0 $ 
there occur realizations such that the effective Hamiltonian $ H(0) $ has 
$ n_0 + 1 $ eigenvalues strictly smaller than $ \varepsilon  $.
Thanks to the non-randomness of each $ \beta_n $, this rules out a flat energy band of $ H $ at $ \varepsilon $. 
Otherwise the number 
of eigenvalues of $ H(0) $ below $ \varepsilon $ would be uniformly bounded in the randomness. 
By a similar argument the almost-sure (purely absolutely continuous) spectrum of $ H $ is seen to coincide with 
the entire positive half-line.\\

\noindent {{\it Dynamic properties.}~
As suggested by the (quasi{-)}classical picture, the non-existence of flat energy bands as supplied by the theorem
should come with ballistic transport along the $ x_2 $-direction. 
To prepare a precise statement, we temporarily return to a typical realization $ b $ of a general ergodic random process
obeying (\ref{eq:mean}). Then (\ref{eq:decomp}) and (\ref{eq:redHs}) imply that 
any normalized 
wave packet $ | \psi_0 \rangle $ in $ {\rm L}^2(\mathbb{R}^2) $ with almost surely finite (time-invariant) kinetic energy, 
  $ \langle \psi_0 |\, H \,| \psi_0 \rangle < \infty $, and (initial) localization along the $ x_2 $-direction in the sense that 
  $ \langle \psi_0 |\, Q_2^2 \, | \psi_0 \rangle < \infty $, has an asymptotic velocity in the sense that 
  the following (strong) long-time-limit relation holds\footnote{The rigorous derivation of (\ref{eq:ball}) is based on the integral form of the Heisenberg equation of motion 
  for $ e^{i t H} Q_2 e^{-itH} $. It is similar to that of the corresponding statement for motion in a periodic scalar potential in \cite{AsKn98}. For details see \cite{LWW04}.}
  \begin{equation}\label{eq:ball}
    \lim_{t \to \infty} \;  t^{-1} \; {\rm e}^{itH}\, Q_2 \, {\rm e}^{-itH} \, | \psi_0 \rangle 
    = V_{2,\infty} \, | \psi_0 \rangle.   
  \end{equation}
Here the (random) asymptotic velocity operator 
\begin{equation}
  V_{2,\infty} := \int_{-\infty}^\infty dk \,  V_{2,\infty}(k) \otimes | k \rangle \langle k | 
\end{equation}
on $ {\rm L}^2(\mathbb{R}^2) $ is related to the derivatives of the energy-band functions similarly as in the quantum theory of single electrons 
in perfect crystals (without external fields) \cite{Bl62},   
\begin{equation}
  V_{2,\infty}(k) := \sum_{n=0}^\infty \frac{d \varepsilon_n(k)}{d k} \, | \varphi_n(k) \rangle \langle \varphi_n(k) |, 
        \quad k \in \mathbb{R}. 
\end{equation}
If the energy band $ \beta_n $ is not flat, $ | \beta_n | > 0 $, 
the (random) group velocity $ d \varepsilon_n(k)/d k $ vanishes at most at countably many 
$ k \in \mathbb{R} $, because $ \varepsilon_n(k) $ is an analytic function of $ k $, almost surely. Moreover, by the Feynman-Hellmann theorem, the positivity of the quantum-mechanical variance and the strict inequality 
$\langle \varphi_n(k) | \, P_1^2\, | \varphi_n(k) \rangle > 0 $, 
Eqs.~(\ref{eq:redH}) and (\ref{eq:redHs}) give the 
upper estimate
$\big(d \varepsilon_n(k)/d k \big)^2 < 2 \varepsilon_n(k) $ (cf. \cite{MaPu97}). 
Taken together, this proves the 

\vspace*{0.1cm}
\noindent
{{COROLLARY.}~~{\it Under the assumptions of the theorem the particle's motion along the $ x_2 $-direction is ballistic in the sense that
  (\ref{eq:ball}) holds with
  $ 0 < \langle \psi_0 | \, V_{2,\infty}^{\, 2} \, | \psi_0 \rangle < 2 \, \langle \psi_0 | \, H \, | \psi_0 \rangle < \infty $, almost surely.}

\vspace*{0.1cm}
In contrast, the particle's motion along the $ x_1 $-direction is bounded. 
Indeed, for a typical realization $ b $ of a 
general ergodic random process obeying (\ref{eq:mean}) the quadratic confinement of the particle along the $ x_1 $-direction
for large $ | x_1 | $ (cf.\ (\ref{eq:erg})) 
implies that any normalized wave packet $ | \psi_0 \rangle $ in $ {\rm L}^2(\mathbb{R}^2) $ 
with almost surely finite kinetic energy and (initial) localization along the $ x_1 $-direction 
in the sense that $ \langle \psi_0 |\, Q_1^2 \, | \psi_0 \rangle < \infty $, 
remains localized in the course of time,
\begin{equation}
  \sup_{t \in \mathbb{R}} \;\; \langle \psi_0 | \, {\rm e}^{itH} \, Q_1^2 \, {\rm e}^{-itH} \, | \psi_0 \rangle < \infty.
\end{equation}

\noindent {{\it Concluding remarks.}~Bounds on the Lifshits tail, that is, 
on the low-energy asymptotics of the integrated density of states of (\ref{def:H}) have been derived in \cite{Uek00} 
under assumptions similar to those of the theorem (but allowing for
$ \overline{b(0)} = 0 $).

For further details, complete proofs, and non-Gaussian RMF's obeying (\ref{eq:mean}) and (\ref{eq:topol}) and 
hence yielding almost surely purely continuous energy spectrum 
and ballistic transport along the $ x_2 $-direction, we refer to \cite{LWW04}.

\ack
We are indebted to Ludwig Schweitzer (Braunschweig, Germany) for hints to the literature. 
This work was partially supported by the Deutsche Forschungsgemeinschaft (DFG) under grant nos. Le 330/12 and Wa 1699/1.

\end{document}